\documentclass[aps,twocolumn,nofootinbib,showpacs,floatfix]{revtex4}
\usepackage{amsmath,psfig}
\usepackage{amsfonts}

\begin{document}
\frenchspacing \title{Critical transient in the Barab\'asi model of
human dynamics} \author{A. Gabrielli, G. Caldarelli} \affiliation{SMC,
INFM-CNR, Dipartimento di Fisica, Univ. ``La Sapienza'',
Piazzale A. Moro 2, 00185-Rome, Italy\\ and Istituto dei Sistemi Complessi
CNR, via dei Taurini 19, 00185-Rome, Italy}

\begin{abstract}
We introduce an exact probabilistic description for $L=2$ 
of the Barab\'asi model for the dynamics of a list of $L$ tasks. 
This permits to study the problem out of stationarity, and to solve 
explicitly the extremal limit case where a critical behavior for 
the waiting time distribution is observed. This behavior deviates at 
any finite time from that of the stationary state.
We study also the characteristic relaxation time for finite time deviations
from stationarity in all cases showing that it diverges in the extremal 
limit confirming that this deviations are important at all time.
\end{abstract}
\maketitle

Queueing theory is a very important branch of probability with
fundamental applications in the study of different human dynamics
\cite{Que1}. Its capability in explaining and modeling complex
behaviors of such activities has potentially important economical
consequences. An example is the prediction
and organization of ``queues'' in hi-tech communications.  Queue stochastic
dynamics are traditionally modeled as homogeneous Poisson processes
\cite{Que2}.  This means that the probability to do a specific
action in the time $dt$ is given by $qdt$, where $q$ is the overall
activity per unit time, and there is no statistical correlation
between non overlapping time intervals.  As a result, the time
interval between different events is predicted to be exponentially
distributed.  Actually, different experimental analysis
\cite{Barabasi05,VODGKB06} have shown that for various human
activities the distribution of waiting times is better fitted by heavy
tail distributions like the power laws characteristic of Pareto
processes.  In order to reproduce such a behavior Barab\'asi
\cite{Barabasi05} has introduced a simple model for a task list with a
heavy tail distribution of waiting times in its stationary state and in
a particular extremal limit.  In this paper we analyze the same model
out of the stationary state introducing an exact step-by-step method
of analysis through which we find the exact waiting time distribution
in the same extremal limit. We show that in this limit the stationary
state is reached so slowly that almost all the dynamics is described
by the transient.

In the Barab\'asi model the list consists at any time in a constant
number $L$ of tasks to execute. Each task has a random priority index
$x_i$ ($i=1,...L$) extracted by a probability density
function (PDF) $\rho(x)$ independently one of each other.  The
dynamical rule is the following: with probability $0\le p\le 1$ the most urgent
task (i.e. the task with the highest priority) is selected, while with
complementary probability $(1-p)$ the selection of the task is done completely at
random. The selected task is executed and removed from the list; it is
then replaced by a fresh new task with random priority extracted again
from $\rho(x)$.  For $p=0$ the selection of the task at each time step
is completely random.  Therefore the waiting time distribution for a
given task is $P(\tau)=(1/L)(1-1/L)^\tau$ and decays exponentially
fast to zero with time constant $\tau_0\simeq L$.  For $p=1$ instead
the dynamics is deterministic selecting at each time step the task in
the list with the highest priority. Due to this extremal nature
the statistics of the dynamics for $p=1$ does not depend on $\rho(x)$.  It
has been shown \cite{Barabasi05,Vazquez05} that the dynamics for
$0<p<1$ reaches a stationary state characterized by a waiting time
distribution $P(\tau)$ which is proportional to $\tau^{-1}$ with a
$p-$dependent upper cut-off and amplitude. For $p\rightarrow 1$ the cut-off
diverges, but the amplitude vanishes, so that in this limit from one
side criticality (i.e. divergence of the characteristic waiting time
$\tau_0$) is approached, but from the other $P(\tau)$ looses sense due
to the vanishing of its amplitude.  From simulations this behavior
does not look to depend on $L$, and the exact analytic solution for
the stationary state has been given for $L=2$ in \cite{Vazquez05}.

Here we propose a different approach to the problem for $L=2$ able to
give a complete description of the task list dynamics also out of the
stationary state. In this way we find the exact waiting time
distribution for $p=1$ which is characterized by power law tails, but
with a different exponent from that found in \cite{Vazquez05} for the
stationary state.  We also show that in this extremal limit the finite
time deviation from the trivial stationary state has diverging
relaxation times. Hence the waiting time distribution at any time is
completely determined by these deviations and differs from the stationary one.

\begin{figure}
\centerline{\psfig{figure=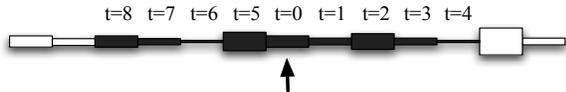,width=8cm}}
\caption{The figure represents the first eight time steps of IP
dynamics in $d=1$. At each time step the throat of the growth
interface with the maximal capillarity (i.e. minimal diameter) is
occupied by the invader fluid. The throats have random capillarities,
i.e.  random diameters and, due to the extremal nature of the
dynamics, its statistical properties do not depend on the PDF from
which capillarities are extracted.
\label{fig1}}
\end{figure}

\begin{figure}
\centerline{\psfig{figure=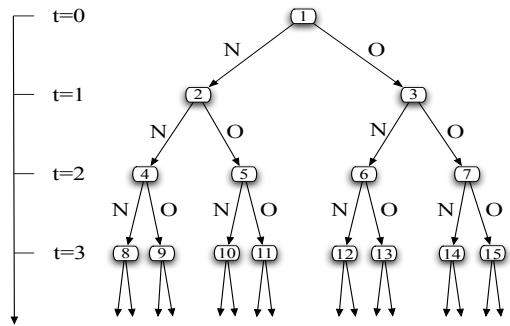,width=7cm,height=4.5cm}}
\caption{Tree representation of all the possible selection paths
(i.e. realizations of the dynamics). At each time step either the fresh new
task ($N$) either the old one ($O$) is selected. In general for $p>0$
$p_O(x,t)$ at the end of a path depend on the specific path. 
Only for $p=1$, when information on priorities and memory are maximal,
$p_O(x,t)$ is the same for all paths.
\label{fig2}}
\end{figure}

The method we use here is called Run Time
Statistics (RTS) \cite{Marsili} and it has been introduced originally
to study an important model of Self Organized Criticality (SOC): the
Invasion Percolation (IP) in dimension $d=2$ \cite{WW83,Ip2} and
related models \cite{BS-prl}.  For $p=1$ this task list model
can be exactly mapped in Invasion Percolation (IP) in $d=1$. The latter has 
this formulation: consider a linear chain
of throats (see Fig.\ref{fig1}), each of which is given, independently of the
others, a random number $x_i$ extracted from the PDF $\rho(x)$ and
representing its capillarity (which is inversely proportional to the
diameter). 
At $t=0$ the invader fluid is supposed to occupy only the
central throat. At each time step the growth interface is given by the
two non occupied throats in contact with the invaded region. The
invaded region grows by occupying the interface
throat with the maximal capillarity (i.e. minimal diameter).  In this
way the interface is updated by eliminating from it the selected throat and including
the next one in the direction of growth.  This is
exactly equivalent to the task list problem for $p=1$ and $L=2$, 
with the set of
the already executed tasks given by the invaded region, and the task
list given by the two throats of interface.  Analogously the case of
the task list with $L>2$ can be mapped for $p=1$ into the case of
IP on a star-like set of throats with $L$ linear
branches \cite{GRC06}.  For $0<p<1$ the task list problem would
correspond to a sort of IP at finite temperature \cite{Ip1} which
introduces a source of noise in the problem permitting random non
extremal growths. The quantity $(1-p)$ is a measure of such thermal
noise which vanishes for $p=1$ and becomes maximal for $p=0$.
In general for growth
processes with quenched disorder it is very difficult to
factorize the statistical weight of a growth path (i.e. selection
sequence) into the product of probabilities of the composing
elementary steps.  The problem is that these single-step probabilities
depend upon all the past history of the process, i.e, dynamics in
quenched disorder present usually strong memory effects.  This feature
makes very difficult to evaluate the statistical weight of any
possible history of the process which is usually the elementary
ingredient to perform averages over the possible realizations of the
dynamics. To overcome this difficulty, RTS gives a
step-by-step procedure to write the exact evolution of the
probabilities of the single steps conditional to the past history and
the related conditional PDF of the random variables attached to the
growing elements (``task priorities'' in this case).  Given two tasks
with respective priorities $x$ and $y$, we call $\eta(x,y)$ the
probability conditional to these values to execute the task with
priority $x$.  The form of $\eta$ is determined by the selection
(i.e. growth in IP) rule of the model. Given the definition of the
Bar\'basi model and being $\theta(x)$ the Heaviside step function, we have:
\begin{equation}
\eta(x,y)= p\theta(x-y) + \frac {1-p} 2\,.
\label{eta}
\end{equation}
If we now suppose that at the $t^{th}$ time-step of the selection dynamics the
variables $x$ and $y$ are statistically independent and have
respectively the PDFs conditional to the past history of the process
$p(x,t)$ and $q(y,t)$, we can write the probability of selecting
the task with priority $x$ conditional to the past history as
\begin{equation}
\mu(t)=\int_0^1\int_0^1 dx \, dy\,p(x,t) q(y,t) \eta(x,y).
\label{eq1} 
\end{equation}
When the selection is done the selected task is removed from the list
and replaced by a fresh new task with a random priority extracted from
$\rho(x)$. The other task remains instead in the list.  We indicate
the first fresh task with $N$ (as ``new"). The PDF of its priority
conditional to the past history is simply $\rho(x)$ as it is new. The
second task is consequently indicated with $O$ (as ``old"), and we
call the PDF of its priority, conditional to the past history of the
task list, as $p_O(x,t)$.  For all $t>0$ and $p>0$ it differs from
$\rho(x)$. Given our particular form (\ref{eta}) of $\eta(x,y)$, $p_O(x,t)$ is,
for $p>0$, more concentrated on the small values of $x$ than
$\rho(x)$. At $t=0$ the initial condition is $p_O(x,0)=\rho(x)$.
Since at each time-step the task $N$ is new, the priorities $x$ and
$y$ of $N$ and $O$ are statistically independent and their joint probability
factorizes into the product $\rho(x)p_O(y,t)$.  Any realization
of the task list (i.e. a selection path) lasting $\tau$ steps can be
represented as a time ordered string of $\tau$ letters $N$ and $O$
(e.g. $NONNONO....$).  In particular the statistical weight of the
selection path $NN..N$ composed by $\tau$ subsequent events $N$ gives
the probability that the waiting time of the task $O$, from the
beginning of the list dynamics, is at least $\tau$. In terms of IP in
$d=1$ this path represents a growth avalanche in one single direction
starting at $t=0$ and lasting at least $\tau$ steps.
We can rewrite Eq.~(\ref{eq1}) for both the cases in
which the task $N$ or $O$ are selected at time $t$:
\begin{equation}
\label{eq1-b}
\left\{
\begin{array}{l}
\mu_N(t)=\int_0^1\int_0^1 dx \, dy\,\rho(x) p_O(y,t) \eta(x,y)\\ \\
\mu_0(t)=1-\mu_N(t)=\int_0^1\int_0^1 dx \, dy\,p_O(x,t)\rho(y) 
\eta(x,y)
\end{array}
\right.
\end{equation}
For each of this two selection events we update consequently the
conditional PDFs of the priorities including this last step in the
past history conditioning probabilities.  As explained above, 
the conditional PDF of the new task replacing the selected one is
$\rho(x)$.  Instead the conditional PDF $p_O(x,t+1)$ at
time $t+1$ of the just unselected task, still in the list, is different in
the two cases above.  If the task $N$ is selected, the task $O$
remains $O$ also at the next time-step (see Fig.\ref{fig2}).  
We can use the
rules of the conditional probability to include the last selection
step in the ``memory" of the past history:
\begin{equation}
\label{eq2}
p_O(x,t+1)=\frac 1 {\mu_N(t)} p_O(x,t) \int_0^1 dy \rho(y)\eta(y,x)\,. 
\end{equation}
If instead $O$ is selected at time $t$, it is removed, and
the $N$ at time $t$ becomes the task $O$ at time $(t+1)$:
\begin{equation}
\label{eq2-b}
p_O(x,t+1)=\frac 1 {\mu_O(t)} \rho(x) \int_0^1 dy p_O(y,t)\eta(y,x)\,. 
\end{equation}
The whole set of all possible selection paths can be
represented as a non-Markovian binary branching process whose realizations
tree is represented in Fig.2.  The initial node (top vertex)
represents the initial situation where one has two tasks with
completely random priorities [i.e. distributed as $\rho(x)$].  
From each node there is a
bifurcation of possible choices: either
task $N$ or $O$ is selected.  Therefore each node of the tree
represents the task list at the end of the time ordered selection path
connecting directly the top vertex with the given node and is
characterized by path-dependent conditional PDF $p_O$ and
probabilities for the next bifurcation $\mu_N$ and $\mu_O$.  The exact
statistical weight of each path on the tree and the conditional PDF $p_O$ at the
end of it can be calculated by applying the above RTS step-by-step
procedure.  Therefore the RTS provides a complete
mathematical description of the task list dynamics. 
Note that the dynamics of the task list
in this model is a binary branching process with
memory, in the sense that the probabilities of a configuration at a
given time depends for $p>0$ on all the past history of the list.  This
memory effect is maximized for $p=1$ when the list dynamics becomes
deterministic and extremal.

A very important quantity in this class of dynamics is the
average priority ``histogram'' \cite{Ip2,Ip1}, that is the
statistical distribution of the priorities of the task list at a given
time $t$ averaged over all the selection paths:
$h(x,t)=[\rho(x)+\left<p_O(x,t)\right>_t]/2$.  
Hence, the evolution of $h(x,t)$ is directly given by
that of $\rho_1(x,t)=\left<p_O(x,t)\right>_t$.  
The equation for its time evolution can be found by observing that 
at each binary branching starting from a node at time $t$ of 
the tree we can say that with probability $\mu_N(t)$ the priority 
conditional PDF $p_O(x,t)$ updates as in Eq.~(\ref{eq2}) and with 
probability $\mu_O(t)$ as in Eq.~(\ref{eq2-b}), i.e.
\begin{eqnarray}
\pi_O(x,t+1;t)&=&p_O(x,t)\int_0^1 dy \rho(y)\eta(y,x)\nonumber\\
&+&\rho(x) \int_0^1 dy p_O(y,t)\eta(y,x)\,,
\nonumber
\end{eqnarray}
where $\pi_O(x,t+1;t)$ is the conditional PDF of the priority
of the task $O$ at time $(t+1)$ conditional to the history only
up to time $t$.
By applying this average from the first time step it is simple to 
show that:
\begin{eqnarray}
\rho_1(x,t+1)&=&\rho_1(x,t)\int_0^1 dy \rho(y)\eta(y,x)\nonumber\\
&+&\rho(x) \int_0^1 dy \rho_1(y,t)\eta(y,x)\,.
\label{vaz}
\end{eqnarray}

This is exactly the basic equation used in \cite{Vazquez05} to study
the task list dynamics in the stationary state, i.e., when
$\rho_1(x,t+1)=\rho_1(x,t)$.  We show in the following that however in
the limit $p\rightarrow 1$ the stationary state is reached only very slowly 
and that the waiting time distribution is determined by the finite time 
deviation from the stationary state at all time.
This waiting time distribution is again a power law, 
but with a different exponent with respect to that found
in \cite{Vazquez05}.  
We now solve exactly the RTS for the case $p=1$.  For our calculation
we use here $\rho(x)=1$ for $x\in[0,1]$ as in \cite{Vazquez05} because the
path statistics, as aforementioned, does not depend on $\rho(x)$ for
$p=1$.  Hence Eqs. (\ref{eq1-b}) above takes the simple form:
\[
\left\{
\begin{array}{l}
\mu^{e}_N(t)
=\int_0^1dx\,(1-x)p^{e}_O(x,t)
\\ \\
\mu^{e}_O(t)=1-\mu^{e}_N(t)=\int_0^1dx\,x p^{e}_O(x,t)\,,
\end{array}
\right.
\]
where the superscript ``$e$'' stands for ``extremal''.  Analogously
Eqs. (\ref{eq2}) and (\ref{eq2-b}) for $p=1$ read respectively
\begin{eqnarray}
&&p^{e}_0(x,t+1)
=\frac{1}{\mu^{e}_N(t)} (1-x)p^{e}_O(x,t) \nonumber\\
&&p^{e}_0(x,t+1)= 
\frac{1}{\mu^{e}_O(t)}\int_x^1dx\,p^{e}_O(x,t)\,.\nonumber
\end{eqnarray}
using $p^{e}_O(x,0)=\rho(x)$ in the above equations, one finds
that, for any selection path, $\mu_N^{e},\mu_O^{e}$ and $p^e_O$
becomes:
\begin{eqnarray}
\label{eq3}
&&\mu_N^{e}(t)=\frac{t+1}{t+2}\;\;\;;\;\;\;
\mu_O^{e}(t)=\frac{1}{t+2}\\
&&p^e_O(x,t)=(t+1)(1-x)^t\,.
\label{eq4}
\end{eqnarray}
Since $p^e_O(x,t)$ is independent of the considered path,
$\rho_1(x,t)$ for $p=1$ coincides with it.  Note that
$\mu_N^{e}(t\!\!\rightarrow\!\!\infty)\rightarrow\! 1$ and
$\rho_1(x,t)=p^e_O(x,t\!\!\rightarrow\!\!\infty)\rightarrow\!\delta(x-0^+)$
[where $\delta(x)$ is the Dirac delta function], i.e., in the
infinite time limit the new fresh task is always selected as the old
one has vanishing priority with probability one.  The fact that both
the $\mu^e$'s and the $p^{e}$'s at time $t$ are the same for each
selection path of length $t$ is a feature of the $p=1$ case.
This is not the case for $0<p<1$
where instead the conditional selection probability and priority PDFs
at time $t$ depend on which specific selection path is considered.  
We now analyze the consequences of Eq.~(\ref{eq3}). It permits to find
the waiting time distribution of a given task entered the list at time
$t_0$. From Eq.~(\ref{eq3}) we find that for $p=1$ the waiting time 
is $\tau=0$ with probability $P(\tau=0;t_0)=(t_0+1)/(t_0+2)$.
The probability that it is still waiting after
$\tau\ge 1$ steps, i.e. at time $t_0+\tau$, is:
\[
W(\tau,t_0)=\frac{1}{t_0+2}\prod_{t'=1}^{\tau-1}\frac{t_0+t'+1}{t_0+t'+2}
=\frac{1}{t_0+\tau+1}\,,
\]
which is the probability of the path $ONN...N$ with
one $O$ event at $t_0$ and $(\tau-1)$ subsequent $N$ events.  Hence the
probability that the waiting time is exactly $\tau\ge 1$ is
\begin{equation}
P(\tau;t_0)=\frac{W(\tau;t_0)}{t_0+\tau+2}
=\frac{1}{(t_0+\tau+1)(t_0+\tau+2)}\,,
\label{eq6}
\end{equation}
which is the statistical weight of the selection path $ONN...NO$ with
one $O$ event at $t_0$, $(\tau-1)$ subsequent $N$ events and a final
$O$ event. Note that this corresponds in IP to the statistical weight
of an avalanche starting a time $t_0$ and lasting $\tau$ steps.  The
probability $P(\tau;t_0)$ decreases as $\tau^{-2}$ for $\tau\gg t_0$ (a behavior 
confirmed by numerical simulations). 
Therefore the waiting time distribution for a task entered at time
$t_0$ is normalizable in $\tau$, but with diverging mean value. This
behavior is different from the power law $P(\tau)\sim \tau^{-1}$
found in \cite{Vazquez05} for the stationary state for $0<p<1$, which
however disappears for $p\rightarrow 1^-$ as its amplitude vanishes in
this limit.  For the opposite limit $t_0\gg \tau\gg 1$ one can write
$P(\tau;t_0)\simeq t_0^{-2}(1-2\tau/t_0)$, which decreases as
$t_0^{-2}$ with $t_0$. This gives the rate of approach in the initial
time $t_0$ to the trivial stationary state
$P(\tau;t_0\rightarrow\infty)=1$ or $0$ respectively if $\tau=0$ or
$\tau\ge 1$. This rate is very slow and there is no
characteristic time after which one can say that the stationary state
is attained in terms of $\tau$ dependence.

In order to study more in detail the approach to the stationary state
for all $p$ and for $p\rightarrow 1$, we analyze Eq.~(\ref{vaz}) out
of the stationary state. First of all we rewrite this equation using
the explicit form (\ref{eta}) of $\eta(x,y)$ for this model and
$\rho(x)=1$ for $x\in [0,1]$:
\begin{eqnarray}
\rho_1(x,t+1)&=&\rho_1(x,t)\left[p(1-x)+\frac{1-p}{2}\right]\nonumber\\
&+&p\int_x^1 dy \rho_1(y,t)+\frac{1-p}{2}\,.
\label{vaz2}
\end{eqnarray}
We now put 
$\rho_1(x,t)=\rho_1^{(s)}(x)+\delta \rho_1(x,t)\,,$
where $\rho_1^{(s)}(x)$ is the stationary solution found in Eq.~(4) of \cite{Vazquez05}:
\begin{equation}
\rho_1^{(s)}(x)=\frac{1+p}{1-p}\frac{1}{[1+\frac{2p}{1-p}x]^2}
\label{stationary}
\end{equation} 
 and
$\delta \rho_1(x,t)$ is the finite time deviation from it.  
For 
$p\rightarrow 1$ the PDF $\rho_1^{(s)}(x)\rightarrow\delta(x-0^+)$ 
and it coincides with Eq.~(\ref{eq4}) for
$t\rightarrow \infty$. Since $\rho_1(x,t)$ and $\rho_1^{(s)}(x)$
are both normalized to unity, we have $\int_0^1 dx\delta
\rho_1(x,t)=0$. Therefore as a first order approximation we  put
$\int_x^1dy\,\delta\rho_1(y,t)\simeq -x\delta\rho_1(x,t)$. 
Taking also the continuous time approximation
$[\delta\rho_1(x,t+1)- \delta\rho_1(x,t)]\simeq
d\delta\rho_1(x,t)/dt$, we can rewrite Eq.~(\ref{vaz2}) in terms 
of $\delta\rho_1(x,t)$ as
\begin{equation}
\frac{d\delta\rho_1(x,t)}{dt}\simeq
-\left(\frac{1-p}{2}+2px\right)\delta\rho_1(x,t)\,.
\label{vaz3}
\end{equation}
Hence $\delta\rho_1(x,t)$ decays exponentially in
time with an $x-$dependent time constant inversely proportional to
$[(1-p)/2+2px]$.  For $p<1$, at each $x$ the perturbation decays
exponentially fast and the stationary state is attained, while for
$p\rightarrow 1$ the time constants becomes proportional to $1/x$
and the perturbation in the region around $x=0$ relaxes very slowly.
But from Eq.~(\ref{stationary}) it is exactly in this region that for
$p\rightarrow 1$ all the measure $\rho^{(s)}(x)$ is concentrated.
This confirms our previous conclusion that for $p\rightarrow 1$ the 
stationary state is very slowly attained and finite time deviation from it
play a fundamental role in determining the rate of decrease of
the waiting time distribution.


In this paper we have studied an interesting queueing model of task
list dynamics introduced by Bar\'abasi. Through a statistical method
called RTS, 
we are able to give a complete probabilistic description of
the dynamics even out of stationarity. We find that for $0<p<1$ finite
time deviations from stationarity relaxes exponentially fast and,
consequently, the dynamics is well described by the stationary
state. However for $p\rightarrow 1$ the stationary state becomes
trivial and finite time deviations relaxes so slowly that the task
list dynamics has to be described as an intrinsically non stationary
dynamics. This is characterized by power law waiting time
distributions with a characteristic exponent which is different from
the one found \cite{Vazquez05} in the stationary state for $0<p<1$.

GC acknowledges support from EU Project DELIS


\begin{thebibliography}{99} 


\bibitem{Que1}
HC Tijms, {\em A First Course in Stochastic Models}, Wiley Chichester, (2003).

\bibitem{Que2}
L. Breuer and D. Baum {\em An Introduction to Queueing Theory and Matrix-Analytic Methods}. 
Springer Verlag (2005).

\bibitem{Barabasi05} A.-L. Barab\'asi, 
Nature (London) {\bf 207}, 435 (2005).

\bibitem{VODGKB06} A. V\'azquez, J. G. Oliveira, Z. Dezs\H{o}, K.-I. Goh, I. Kondor, A.-L. 
Barab\'asi Physical Review E {\bf 73}, 036127 (2006).


\bibitem{Vazquez05} A. Vazquez, Physical Review Letters {\bf 95}, 248701 (2005).

\bibitem{Marsili} M. Marsili, Journal of Statistical Physics,
	{\bf 77} 733--754,(1994); A. Gabrielli, M. Marsili, R. Cafiero, L. Pietronero,
	Journal of Statistical Physics, {\bf 84} 889--893, (1996).

\bibitem{WW83} 
D. Wilkinson and J.F. Willemsen, Journal of Physics A, {\bf 16}, 3365-3376 (1983). 

\bibitem{Ip2} A. Gabrielli, R. Cafiero, M. Marsili, and L. Pietronero,
Phys. Rev. E, {\bf 54}, 1406 (1996).

\bibitem{BS-prl} M. Felici, G. Caldarelli, A. Gabrielli, and L. Pietronero,
Phys. Rev. Lett., {\bf 86}, 1896 (2001).

\bibitem{GRC06}
A. Gabrielli, F. Rao, G. Caldarelli, in preparation.

\bibitem{Ip1} A. Gabrielli, G. Caldarelli, L. Pietronero, 
Physical Review E {\bf 62}, 7638--7641 (2000)


\end{thebibliography}
\end{document}